\documentclass[12pt]{article}

\usepackage{amsmath}
\usepackage{amssymb}

\newcommand{\eps}{\varepsilon}
\newcommand{\sn}{\mathrm{sn}}
\newcommand{\prt}{\partial}
\newcommand{\br}{{\bf r}}
\newcommand{\bu}{{\bf u}}
\newcommand{\la}{\lambda}

\begin{document}

\title{Two-dimensional periodic waves in a supersonic flow of a Bose-Einstein
condensate}

\author{G.A. El$^1$, Yu.G. Gladush$^2$, A.M. Kamchatnov$^2$  \\
$^1$ Department of Mathematical Sciences,\\ Loughborough University,
Loughborough LE11 3TU, UK \\
$^2$Institute of Spectroscopy, Russian Academy of Sciences,\\ Troitsk
142190, Moscow Region, Russia }

\maketitle

\begin{abstract}
Stationary periodic solutions of the two-dimensional
Gross-Pitaevskii equation are obtained and analyzed for different
parameter values in the context of the problem of a supersonic flow
of a Bose-Einstein condensate past an obstacle. The asymptotic
connections with the corresponding periodic solutions of the
Korteweg-de Vries and nonlinear Schr\"odinger equations are studied
and typical spatial wave distributions are discussed.
\end{abstract}


\section{Introduction}

The Gross-Pitaevskii (GP) equation plays a prominent role in the
description of nonlinear dynamics of Bose-Einstein condensates (BEC)
(see, e.g., \cite{ps}). It describes, in the so-called mean-field
approximation, the behavior of the order parameter
$\psi(\mathbf{r})$ (the ``condensate wave function'') and has the
form
\begin{equation}\label{1-1}
   i\hbar\frac{\prt \psi}{\prt t}=-\frac{\hbar^2}{2m}\Delta\psi+
   V(\mathbf{r})\psi+g|\psi|^2\psi,
\end{equation}
where $V(\mathbf{r})$ denotes the potential of the external forces
acting on the condensate (e.g. the confining trap potential), $g$ is
an effective coupling constant arising due to inter-atomic
collisions with the $s$-wave scattering length $a_s$ (positive for
repulsive interactions), $g=4\pi\hbar^2a_s/m,$ $m$ being the atomic
mass. The GP equation (\ref{1-1}) takes into account the dispersive
and nonlinear properties of the condensate which can give rise to
various nonlinear structures in a BEC flow. In particular, vortices,
bright (for $g<0$) and dark (for $g>0$) solitons have been
extensively studied theoretically in the framework of the GP
equation  and observed in experiments (see, e.g., \cite{ps} and
references therein).

An important insight into the structure of the solutions of the GP
equation is provided by the fact that its potential-free,
one-dimensional reduction coincides with the integrable nonlinear
Schr\"odinger equation
\begin{equation}\label{nls}
  i\psi_\tau +\psi_{XX} -  2\sigma |\psi|^2\psi=0 \, .
\end{equation}
Here $X=x/\xi$, $\tau = t c_s/\xi $ and $\sigma= \hbox{sgn}\, g$,
where $\xi=\hbar/\sqrt{2mn_0g}$ is the healing length and
$c_s=\hbar/\sqrt{2}m\xi$ is the sound velocity in a BEC of density
$n_0$.

In recent experiments \cite{cornell05} on free expansion of a
non-rotating BEC after its release from a trap, the new interesting
blast wave patterns have been observed. Using the modulation
solutions of the one-dimensional NLS equation (\ref{nls})
(defocusing case), obtained earlier in \cite{gk87, eggk95, kod99,
kku}, and numerical simulations in 2D and 3D, these blast waves have
been identified in \cite{kgk, damski, hoefer} with expanding
dispersive shock waves, which represent oscillatory counterparts of
classical evolutionary gas-dynamic shocks.

Yet another type of nonlinear wave patterns has been observed in
another series of experiments on the flow of a non-rotating BEC past
macroscopic obstacles also reported in \cite{cornell05}. In
\cite{ek} these structures have been associated with spatial
dispersive shock waves. Spatial dispersive shock waves represent
dispersive analogs of the the well-known viscous spatial shocks
(oblique jumps of compression) occurring in supersonic flows of
compressible fluids past obstacles. In a viscous fluid, the shock
can be represented as a narrow region within which strong
dissipation processes take place and the thermodynamic parameters of
the flow undergo sharp change. On the contrary, if viscosity is
negligibly small compared with dispersion effects, the shock
discontinuity resolves into an expanding in space oscillatory
structure which transforms gradually, as the distance from the
obstacle increases, into a ``fan'' of stationary solitons. The
theory of the generation of such spatial dispersive shock waves has
been developed in \cite{gke} in quite general terms for supersonic
dissipationless flows past a slender body when the flow can be
asymptotically described by the Korteweg-de Vries (KdV) equation and
an effective  description of  dispersive shocks becomes possible via
the Whitham modulation theory \cite{whitham1,whitham2, gurpit}.

In a different approximation, for {\it highly} supersonic flow past
slender body, for a certain range of parameters, the GP equation
{\it asymptotically} reduces to the one-dimensional nonlinear
Schr\"odinger (NLS) equation (\ref{nls}), albeit for a completely
different set of independent variables \cite{ek}.  As a result, the
problem of stationary dispersive shock waves in this case can be
treated by similar methods of Whitham's theory. It is essential that
in all cited papers, the analytical advances in the description of
the dispersive shock waves have become possible owing to a complete
integrability of the KdV and 1D NLS equations resulting in the
possibility to represent the associated modulation Whitham systems
in Riemann diagonal form (see  \cite{kamch2000} and references
therein), which dramatically simplifies further analysis.

However, in real experiments the obstacles  cannot be treated as
slender bodies and the flow is not highly supersonic, so the spatial
dispersive shock waves must be considered in the frame of the {\it
full}, unapproximated GP equation (\ref{1-1}). In such a study, one
inevitably faces the major obstruction to obtaining analytical
solutions to an initial or boundary value problem because the
multi-dimensional GP equation, even in the absence of the external
forcing term, is a non-integrable system and powerful spectral
methods (inverse scattering transform, finite-gap integration) are
not available for it. However, the Whitham modulation approach still
remains a possibility provided a minimal structure (availability of
a certain number of conservation laws and a traveling periodic wave
solution) is present \cite{el05}.

In the framework of the Whitham modulation theory, a dispersive
shock wave is considered as a modulated nonlinear periodic wave
whose parameters change slowly on a scale about one wavelength and
one period (see \cite{whitham2, kamch2000} for instance). For a
certain spatio-temporal domain, the dispersive shock represents a
train of solitons well separated from each other. In the context of
the BEC flow past an obstacle a 2D spatial dispersive shock wave is
asymptotically (far enough from the obstacle) represented as a
``fan'' of spatial dark solitons. When the obstacle is not very
large, only one soliton can be generated. This simplest case has
been considered recently in \cite{egk} where exact spatial soliton
solution of the GP equation (\ref{1-1}) was found for condensate
with a flow and it was shown by a numerical simulation that such
``oblique'' solitons can be generated by a supersonic flow past an
obstacle. To extend this theory to ``multi-soliton'' dispersive
shocks represented by a modulated periodic wave, it is necessary,
first of all, to find stationary periodic solutions of the GP
equation (\ref{1-1}), and this is the main aim of the present paper.
We also note that the periodic solutions of the GP equation could be
important in other, than BEC, areas where the multi-dimensional NLS
equation is used to model nonlinear wave propagation.

\section{Periodic solution}

The stationary solutions of Eq.~(\ref{1-1}) can be sought in the
form
\begin{equation}\label{3-1}
   \psi(\br)=\sqrt{n(\br)}\exp\left(\frac{i}{\hbar}
   \int^{\br}{\bu}(\br')d\br'  \right ) \exp\left(-\frac{i\mu}{\hbar}t\right),
\end{equation}
where $n(\br)$ is the density of atoms in the BEC, ${\bf u}(\br)$
denotes its velocity field and $\mu$ is the chemical potential.
It is convenient to introduce the dimensionless variables
\begin{equation}\label{3-2}
   \tilde{\br}=\br/\sqrt{2}\xi,\quad
   \tilde{n}=n/n_0,\quad \tilde{\bf u}={\bf u}/c_s.
\end{equation}
Substituting Eq.~(\ref{3-1}) into
(\ref{1-1}) and separating real and imaginary parts we obtain a
system of equations for the density $n(x,y)$ and the two components of
the velocity field $\bu=(u(x,y),v(x,y))$,
\begin{equation}\label{3-3}
\begin{split}
   (nu)_x+(nv)_y=0,\\
   uu_x+vu_y+n_x+\left(\frac{n_x^2+n_y^2}{8n^2}-
   \frac{n_{xx}+n_{yy}}{4n}\right)_x=0,\\
   uv_x+vv_y+n_y+\left(\frac{n_x^2+n_y^2}{8n^2}-
   \frac{n_{xx}+n_{yy}}{4n}\right)_y=0,
   \end{split}
\end{equation}
where we have omitted tildes for convenience of notation.

One can see that the $(0+2)$ reduction (\ref{3-3}) of the GP
equation (\ref{1-1}) is drastically different from its (1+1) NLS
reduction (\ref{nls}) (which also can be represented in a
hydrodynamic form by the change of variables analogous to
(\ref{3-1}) -- see Section 4). Putting aside the subtle
integrability aspects (the equation (\ref{nls}) is a completely
integrable system while there is no indication of integrability for
the system (\ref{3-3})) we note that, first of all, the scalar
system corresponding to the one-dimensional NLS equation (\ref{nls})
consists of {\it two} equations (for $n$ and $u$) while the spatial
(0+2) case leads to {\it three} equations (\ref{3-3}). As a result,
the structure of the system (\ref{3-3}) is significantly more
complicated. This can already be seen by comparing the
dispersionless limits of (\ref{nls}) and (\ref{3-3}). Indeed, the
dispersionless limit one-dimensional NLS equation (\ref{nls})
coincides with the classical shallow-water equations (or classical
gas-dynamic equations with the adiabatic index $\gamma=2$) which are
always hyperbolic. Contrastingly, the dispersionless limit of
(\ref{3-3}) coincides with {\it stationary two-dimensional}
gas-dynamic ($\gamma=2$) equations which have a mixed
elliptic-hyperbolic structure depending on the absolute value of the
flow velocity. Thus, one can expect a considerable difference in the
behavior of the periodic solutions and their modulations in these
two different reductions of the GP equation.

We look for the solution of the system (\ref{3-3}) in the form of a
``travelling'' wave
\begin{equation}\label{3-4}
    n=n(\theta),\quad u=u(\theta),\quad v=v(\theta),
\end{equation}
where $\theta=x-ay$, $a$ being the ``slope'' parameter of the
stationary wave (the wave crests lie on parallel lines with the
slope $a$ to $y$-axis). Under this {\it ansatz}, the first equation
(\ref{3-3}) gives at once
\begin{equation}\label{4-1}
    u-av=\frac{A}n,
\end{equation}
where $A$ is the integration constant, and the other two equations
reduce to
\begin{equation}\label{4-2}
    n_\xi^2-2nn_{\xi\xi}+2n^3-\frac{2B}{1+a^2}n^2+\frac{A^2}{1+a^2}=0,
\end{equation}
where $B$ is another integration constant and we have introduced new
independent variable
\begin{equation}\label{4-3}
    \xi=\frac{2\theta}{\sqrt{1+a^2}}=\frac{2(x-ay)}{\sqrt{1+a^2}}.
\end{equation}
One can verify by a direct substitution that equation (\ref{4-2})
has the first integral
\begin{equation}\label{4-4}
    n_\xi^2=n^3-\frac{2B}{1+a^2}n^2-\frac{2C}{1+a^2}n-\frac{A^2}{1+a^2},
\end{equation}
where $C$ is an arbitrary constant.  Equation (\ref{4-4}) has the
well-known solution in terms of elliptic functions. To write it
down, we denote the zeroes of the polynomial in the right hand side
of Eq.~(\ref{4-4}) as $p_1,p_2,p_3$, so that
\begin{equation}\label{4-5}
    n_\xi^2=(n-p_1)(p_2-n)(p_3-n),\quad p_1\leq p_2\leq p_3,
\end{equation}
and suppose that $n=p_1$ at $\xi=0$. As a result, we obtain
\begin{equation}\label{4-6}
    n=p_1+(p_2-p_1)\sn^2\left(\sqrt{p_3-p_1}\,\xi/2;m\right),
\end{equation}
where $\sn(\theta;m)$ is the Jacobi elliptic sine and
\begin{equation}\label{4-7}
    m=\frac{p_2-p_1}{p_3-p_1}
\end{equation}
is the modulus. The constants $A,B,C$ are connected with the zeroes
$p_1,p_2,p_3$ by the relations
\begin{equation}\label{4-8}
    p_1+p_2+p_3=\frac{2B}{1+a^2},\quad p_1p_2+p_1p_3+p_2p_3=-\frac{2C}{1+a^2},
    \quad p_1p_2p_3=\frac{A^2}{1+a^2}.
\end{equation}
It is worth noticing that the components $(u,v)$ of the velocity
field are not determined unambiguously by the constants
$p_1,p_2,p_3$ and $a$. Indeed, if $n$ is known, we have only one
equation (\ref{4-1}) for calculation of $u$ and $v$. Another
equation can be added, if we restrict ourselves to the consideration
of potential flows by imposing the condition
\begin{equation}\label{4-9}
    u_y=v_x,
\end{equation}
which is consistent with the system (\ref{3-3}) (see \cite{egk}),
and for a single-phase wave (\ref{3-4}) yields at once
\begin{equation}\label{4-10}
    au+v=D.
\end{equation}
Here $D$ is an additional integral of `motion' which owes its
existence to the Bernoulli theorem for the system (\ref{3-3}).
Equation (\ref{4-10}) implies that the same spatial periodic profile
of the density (\ref{4-6}) with the slope $a$ can be supported by
different potential velocity fields. If we fix $u$ and $v$ at some
point, then the constant $D$ becomes determined, as well as the
velocity components everywhere. Just this situation occurs in the
case of the soliton solution \cite{egk} where the velocity
components are supposed to be known at $|x|\to\infty$. Indeed,
considering $m=1$ in (\ref{4-6}) and assuming $n =1$, $u = M=
\hbox{constant}$, where $M$ is the Mach number and $v = 0$ as
$|x|\to \infty$ we arrive at the oblique dark soliton solution
obtained in \cite{egk},
\begin{equation}\label{oblique}
n=1-\frac{1-q}{\cosh^2[\sqrt{1-q}\, (x-ay)/(1+a^2) ]},
\end{equation}
where  $q=M^2/(1+a^2)$ and the velocity components are given by
\begin{equation}\label{vel}
    u=\frac{M(1+a^2n)}{(1+a^2)n},\quad v=-\frac{aM(1-n)}{(1+a^2)n}.
\end{equation}
We note that we are not concerned here with the stability of the
obtained periodic solutions which calls for further nonlinear
modulation analysis. The stability of the oblique dark solitons
(\ref{oblique}) for $M>1$ was established in \cite{egk} numerically.

It is important now to investigate the behaviour of the obtained
periodic solution for the parameter values corresponding to some
physically interesting asymptotic reductions of the system
(\ref{3-3}).

\section{Small-amplitude nonlinear periodic waves}

As was indicated in \cite{egk}, if we consider a flow of BEC
corresponding to small deviations from a uniform and homogeneous
supersonic flow with $n=1$, $u=M,$ $v=0$ $(M>1)$, and make
asymptotic expansions
\begin{equation}\label{5-1}
    n=1+\eps n_1+\eps^2n_2+\ldots,\quad u=M+\eps u_2+\eps^2n_2+\ldots,\quad
    v=\eps v_1+\eps^2v_2+\ldots,
\end{equation}
where $\eps\ll 1$ is a small parameter, then substitution of
(\ref{5-1}) into (\ref{3-3}) followed by introduction of the scaled
variables
\begin{equation}\label{5-2}
    \zeta=\eps^{1/2}(x-Vy),\quad \tau=\eps^{3/2}y,
\end{equation}
leads, according to the standard reductive perturbation method, to
relations
\begin{equation}\label{5-3}
    u_1=-\frac{n_1}M,\quad v_1=\frac{V}Mn_1,\quad V=\sqrt{M^2-1},
\end{equation}
where $n_1$ obeys the KdV equation
\begin{equation}\label{5-4}
      n_{1,\tau}-\frac{3M^2}{2\sqrt{M^2-1}}n_1  n_{1,\zeta}+
    \frac{M^4}{8\sqrt{M^2-1}}n_{1,\zeta\zeta\zeta}=0.
\end{equation}
Its periodic solution is well known (see, e.g. \cite{kamch2000}) so
that the density profile can be expressed after returning to
original $(x,y)$-coordinates as
\begin{equation}\label{5-5}
    n=1-\frac12M^2\eps(\la_3-\la_1-\la_2)+M^2\eps(\la_3-\la_2)
    \sn^2\left[\sqrt{\la_3-\la_1}\,\eps^{1/2}(x-ay),m\right],
\end{equation}
where
\begin{equation}\label{5-6}
    a=\sqrt{M^2-1}-\frac{\eps(\la_1+\la_2+\la_3)M^2}{4\sqrt{M^2-1}},\quad
    m=\frac{\la_3-\la_2}{\la_3-\la_1},
\end{equation}
and $\la_1\leq\la_2\leq\la_3$ are the parameters arising in the
finite gap integration method of the KdV equation (see
\cite{kamch2000}).

It is instructive to establish a direct asymptotic correspondence
between fully nonlinear periodic solution (\ref{4-6}) of the GP
equation and its small-amplitude KdV counterpart (\ref{5-5}). For
that, we first represent the arbitrary parameters $p_1, p_2, p_3$ in
the form of asymptotic expansions in the small parameter $\eps$
\begin{equation}\label{as}
p_i=1+\eps p_i^{(1)} + \dots \, , \quad i=1,2,3\, ,
\end{equation}
and then substitute the expansions (\ref{5-1}), (\ref{5-3}),
(\ref{as}) into the periodic solution (\ref{4-6}) to obtain the same
solution (\ref{5-5}) but parameterised by $p_i^{(1)}$ so that
comparison with (\ref{5-5}) yields
\begin{equation}\label{6-3}
    \begin{split}
    &p_1=1-\frac12M^2\eps(\la_3-\la_1-\la_2),\\
    &p_2=1-\frac12M^2\eps(\la_2-\la_1-\la_3),\\
        &p_3=1-\frac12M^2\eps(\la_1-\la_2-\la_3),
    \end{split}
\end{equation}
Then, using analogous asymptotic expansion for $a$,
\begin{equation}\label{aas}
      a=a^{(0)}+\eps a^{(1)}+ \dots\, ,
\end{equation}
the relation (\ref{4-1}) and the last relation in (\ref{4-8}), we
recover the asymptotic expression (\ref{5-6}) for the slope. Inverse
expressions for $\la_j$´s in terms of $p_j$´s, following from
(\ref{6-3}) are,
\begin{equation}\label{6-4}
    \la_1=\frac{p_1+p_2-2}{M^2\eps},\quad \la_2=\frac{p_1+p_3-2}{M^2\eps},\quad
    \la_3=\frac{p_2+p_3-2}{M^2\eps}.
\end{equation}

The soliton solution of the KdV equation corresponds to $m=1$ in
(\ref{5-5}). In terms of the original parameters $p_j$, we have
$p_2=p_3=1$ (see (\ref{4-7}), (\ref{6-3})),  which implies by
(\ref{6-4}) that $\la_1=\la_2 \equiv \la$ and $\la_3=0$. Then
\begin{equation}\label{7a-1}
    p_1=1+M^2\eps\la
\end{equation}
and, since $p_1 \le p_2 \le p_3$ (see (\ref{4-5})), we have $p_1 \le
1$ and thus $\la \le 0$. Now, from (\ref{5-5}) we have for $m=1$ the
small-amplitude dark soliton profile
\begin{equation}\label{7a-2}
    n=1-\frac{-M^2\eps\la}{\cosh^2[\sqrt{-\eps\la}(x-ay)]},
\end{equation}
and from (\ref{5-6}) its slope is
\begin{equation}\label{assa}
a=\sqrt{M^2-1} - \frac{\eps \la M^2}{2\sqrt{M^2-1}}\, .
\end{equation}
Since $\la \le 0$, one can see from (\ref{assa}) that $a \ge a_M$,
where $a_M=\sqrt{M^2-1}$, i.e. the shallow (KdV) dark solitons
always lie within (and close to) the Mach cone.

If we introduce the inverse half-width $\kappa$ of the soliton
according to
\begin{equation}\label{7a-3}
    \kappa=2\sqrt{-\eps\la},
\end{equation}
then (\ref{7a-2}) assumes a more conventional form
\begin{equation}\label{7a-4}
    n=1-\frac{M^2\kappa^2}{4\cosh^2[\kappa(x-ay)/2]},
\end{equation}
which also follows directly from the oblique dark soliton solution
(\ref{oblique}) in the small-amplitude limit \cite{egk}.

Thus, we have established an asymptotic correspondence between the
stationary periodic two-dimensional solution of the GP equation
characterised by four independent parameters and the familiar
three-parameter cnoidal wave solution of the KdV equation.

\section{Periodic stationary waves with large slopes in highly supersonic flow}

If the flow is highly supersonic $(M\gg 1)$, then the system
(\ref{3-3}) can be reduced to the NLS equation \cite{ek}. Indeed,
after introduction of the new variables
\begin{equation}\label{7-1}
   u=M+u_1+O(1/M),\quad T=x/2M,\quad Y=y,
\end{equation}
where $u_1 \to 0$ as $|Y| \to \infty$,   we arrive, to leading order
in $M^{-1}$, at the system
\begin{equation}\label{7-2}
\begin{split}
   \tfrac12 n_T+(nv)_Y=0,\\
   \tfrac12 v_T+vv_Y+n_Y+\left(\frac{n_Y^2}{8n^2}
   -\frac{n_{YY}}{4n}\right)_Y=0,
   \end{split}
\end{equation}
\begin{equation}\label{7-3}
   \tfrac12 u_{1T}+vu_{1Y}=0.
\end{equation}
The leading term of the highly supersonic expansion of the
potentiality condition (\ref{4-9}) together with Eq.~(\ref{7-3})
implies $u_1 \equiv 0$. The decoupled from (\ref{7-3}) equations
(\ref{7-2}) represent the hydrodynamic form of the 1D NLS equation
\begin{equation}\label{7-4}
   i\Psi_T+\Psi_{YY}-2|\Psi|^2\Psi=0
\end{equation}
for a complex field variable
\begin{equation}\label{7-5}
\Psi=\sqrt{n}\exp\left (i\int^Y v(Y',t)dY' \right ).
\end{equation}
Note that this NLS equation, unlike the exact (1+1) reduction
(\ref{nls}) of the GP equation, represents a (0+2) asymptotic {\it
approximation} and, in addition, contains a completely different,
compared to (\ref{nls}), set of independent variables. Periodic
solution of the NLS equation (\ref{7-4}) is well known (see, e.g.,
\cite{kamch2000}) and for the density $n=|\Psi|^2$ can be written in
the form
\begin{equation}\label{7-6}
    n=\nu_1+(\nu_2-\nu_1)\sn^2\left[\sqrt{\nu_3-\nu_1}\left(y-
    \frac{s_1}{2M}x\right),m\right],
\end{equation}
where
\begin{equation}\label{7-7}
    s_1=\la_1+\la_2+\la_3+\la_4,\quad m=
    \frac{(\la_2-\la_1)(\la_4-\la_3)}{(\la_3-\la_1)(\la_4-\la_2)}
\end{equation}
and $\nu_1 \le \nu_2 \le \nu_3$ are expressed in terms of the
parameters $\la_1\leq\la_2\leq\la_3\leq\la_4$ as follows,
\begin{equation}\label{7-8}
    \begin{split}
    \nu_1=\tfrac14(\la_1-\la_2-\la_3+\la_4)^2,\\
    \nu_2=\tfrac14(\la_1-\la_2+\la_3-\la_4)^2,\\
    \nu_3=\tfrac14(\la_1+\la_2-\la_3-\la_4)^2,
    \end{split}
\end{equation}
and $m= (\nu_2-\nu_1)/(\nu_3-\nu_1)$.

For $M\gg1$ the stationary wave has a large slope with respect to
the $y$-axis, that is $a\gg1$ and the general solution (\ref{4-6})
assumes the form
\begin{equation}\label{8-1}
    n=p_1+(p_2-p_1)\sn^2\left(\sqrt{p_3-p_1}(y-x/a); m\right).
\end{equation}
Since the asymptotic solution (\ref{8-1}) of the GP equation is
characterised by four parameters $p_1,p_2,p_3, a$, the
correspondence with the four-parameter ($\nu_1, \nu_2, \nu_3, s_1$)
periodic solution (\ref{7-6}) is readily established by a direct
comparison:
\begin{equation}\label{8-5}
    p_1=\nu_1,\quad p_2=\nu_2, \quad p_3=\nu_3, \quad a =
    \frac{2M}{s_1} \, .
\end{equation}

The dark soliton reduction of the NLS cnoidal wave solution is
obtained by putting $\la_2=\la_3$,  i.e. $m=1$, in (\ref{7-6}). To
get a unit pedestal for the soliton it is convenient to choose
\begin{equation}\label{8-2}
    \la_1=-1,\quad   \la_4=1.
\end{equation}
Then, denoting  $\la_2=\la_3 \equiv \la $ we get from (\ref{8-5}),
(\ref{7-7})
\begin{equation}\label{8-6}
    a=\frac{M}{\la}\, .
\end{equation}
Thus, in the soliton limit we obtain
\begin{equation}\label{9-1}
    n=1-\frac{1-\la^2}{\cosh^2[\sqrt{1-\la^2}(y-x/a)]}
\end{equation}
which agrees with  the asymptotic representation of a highly
supersonic oblique GP soliton in \cite{egk}. Without loss of
generality we consider the waves in the upper half-plane ($a>0$),
then it follows from (\ref{8-2}) that $0 \le \la \le 1$. Thus,
$a>a_M$ and, therefore, the dark solitons in highly supersonic flows
are always generated within the Mach cone.

If the NLS soliton is shallow, then it must be consistent with the
corresponding asymptotic as $M\gg1$ of the KdV soliton. Indeed, we
introduce $\la^2=1-M^2\kappa^2/4$, where $M\kappa\ll1$, and obtain
the KdV soliton solution (\ref{7a-4}) but now with $a$ equal to
\begin{equation}\label{9-2}
    a=M/\sqrt{1-M^2\kappa^2/4}\cong M+\tfrac18M^3\kappa^2
\end{equation}
which is the approximation of Eq.~(\ref{5-6}) with chosen values
of $\la_i$ and $M\gg1$.

The connection between the soliton parameters $\la$ for the KdV and
the ''shallow'' NLS limits is given by the expression
\begin{equation}\label{9-3}
    \la_{NLS}^2=1+M^2\eps\la_{KdV}.
\end{equation}
Since $\la_{KdV}\le 0$, we have $\la_{NLS} \le 1$ but
$1-\la_{NLS}\ll1$.

\section{Linear waves}

At last, let us consider the limit of linear waves propagating on a
constant background (this case is more general than the linear wave
limit within the KdV approximation (\ref{5-4}) as it does not imply
the long-wave scaling (\ref{5-2})). In this case Eq.~(\ref{4-6})
with $m\ll1$ (i.e. $p_2-p_1\ll1$) can be transformed to
\begin{equation}\label{10-1}
    n=1-\tfrac12(p_2-p_1)\cos\left[2\sqrt{\frac{p_3-p_1}{1+a^2}}(x-ay)\right],
\end{equation}
where we have assumed that the mean background density is equal to
unity, $(p_1+p_2)/2=1$.

Further, in the limit of vanishing amplitude $p_2\to p_1$ and large
wavelength we take
\begin{equation}\label{10-2}
    p_1=p_2=1,\quad p_3=1+M^2\eps,
\end{equation}
and obtain for $M^2\eps\ll 1$
\begin{equation}\label{10-3}
    a=\sqrt{M^2-1}-\frac{M^2\eps}{2\sqrt{M^2-1}}.
\end{equation}
Now we have $a<a_M$, which means that the {\it linear} long waves,
in contrast to solitons, are always generated {\it outside} the Mach
cone. If we denote
\begin{equation}\label{10-4}
    2\sqrt{\frac{p_3-p_1}{1+a^2}}\cong2\sqrt{\eps}=K,\quad -\tfrac12(p_2-p_1)=Q,
\end{equation}
then we obtain the stationary linear wave solution in the form
\begin{equation}\label{10-5}
    n=1+Q\cos\left[K\left(x-\left(\sqrt{M^2-1}-\frac{M^2K^2}{8\sqrt{M^2-1}}\right)y\right)
    \right]\, ,
\end{equation}
which maps to the tail of the oblique soliton solution in the KdV
limit (\ref{7a-4}) by the change $\kappa\mapsto iK$.

As was noted above, the general formula (\ref{10-1}) describes
linear waves of an arbitrary wavelength and it can be mapped to the
tails of the general soliton solution (\ref{oblique}).

\section{Conclusions}

We have obtained the family of exact fully nonlinear stationary
periodic solutions of the 2D Gross-Pitaevskii equation and studied
in detail their particular asymptotic reductions corresponding to
the solutions of the KdV and NLS equations.

The obtained solutions provide a basis for further studies connected
with the description of dispersive shock waves observed in recent
experiments \cite{cornell05} of the flow of a BEC past obstacles as
well as in numerical simulations \cite{egk}. Some straightforward
implications about the characteristic features of the wave patterns
arising in the flow of a BEC past an obstacle have been made from
the expressions for the slope $a$ in the obtained asymptotic
reductions of the full periodic solution. The deep solitons
asymptotically described by the NLS equation have large slopes,
while the shallow solitons obey the KdV equation and have slopes
close the Mach cone, $a_M=\sqrt{M^2-1}$, and they are located inside
the Mach cone. The linear wave packets are always located outside
the Mach cone. Detailed theories of these waves patterns generated
by the flow of a BEC past obstacles will be developed elsewhere.

\section*{Acknowledgements}

A.M.K. and Yu.G.G. are grateful to RFBR (grant 05-02-17351) for
financial support. A.M.K. also thanks EPSRC for financial support of
his visit to Loughborough University, where this work has been
initiated.


\begin{thebibliography}{99}

\bibitem{ps} L.P. Pitaevskii and S. Stringari, {\it Bose-Einstein
Condensation,} (Cambridge University Press, Cambridge, 2003).

\bibitem{cornell05} E.A. Cornell, talk at  ``Conference on
Nonlinear Waves, Integrable Systems and their Applications", (Colorado
Springs, June 2005); available at http://jilawww.colorado.edu/bec/papers.html.

\bibitem{gk87}  A.V. Gurevich, A.L. Krylov, Sov. Phys. JETP {\bf 65,} 944 (1987).

\bibitem{eggk95}
G.A. El, V.V. Geogjaev, A.V. Gurevich, and A.L. Krylov, Physica D
{\bf 87}, 186 (1995).

\bibitem{kod99}
Yu. Kodama,  SIAM J. Appl.\ Math.\  {\bf 59}, 2162 (1999).

\bibitem{kku} A.M. Kamchatnov, R.A. Kraenkel, and B.A. Umarov,
Phys. Rev. E {\bf 66,}   036609 (2002).

\bibitem{kgk} A.M. Kamchatnov, A. Gammal, and R.A. Kraenkel,
Phys. Rev. A {\bf 69,} 063605 (2004).

\bibitem{damski} B. Damski, Phys.Rev. A {\bf 69,} 043610 (2004).

\bibitem{hoefer} M.A. Hoefer, M.J. Ablowitz, I. Coddington, E.A. Cornell, P. Engels, and V.
Schweikhard, Phys. Rev. A {\bf 74,} 023623 (2006) .

\bibitem{ek} G.A. El and A.M. Kamchatnov, Phys. Lett A {\bf 350,} 192 (2006);
erratum: Phys. Lett. A {\bf 352,} 554 (2006).

\bibitem{gke} A.V. Gurevich, A.L. Krylov, V.V. Khodorovskii, and
G.A. El, { JETP,} {\bf 81,} 87 (1995); {\bf 82,} 709 (1996).

\bibitem{whitham1} G.B. Whitham,
Proc. Roy. Soc. London, {\bf 283,} 238 (1965).

\bibitem{whitham2} G.B. Whitham, {\it Linear and Nonlinear Waves,} Wiley-Interscience,
New York (1974).

\bibitem{gurpit} A.V. Gurevich and L.P. Pitaevskii, Sov.\ Phys.\ JETP {\bf 38}, 291 (1974).

\bibitem{kamch2000}  A.M. Kamchatnov, {\it Nonlinear Periodic Waves and
Their Modulations,} World Scientific, Singapore (2000).

\bibitem{el05}
G.A. El, Chaos {\bf 15,} 037103 (2005).

\bibitem{egk} G.A. El, A. Gammal, and A.M. Kamchatnov, e-print
nlin.PS/0604044.


\end{thebibliography}
\end{document}